\documentclass[twocolumn,showpacs,superscriptaddress]{revtex4}
\usepackage{graphicx}
\usepackage{natbib}
\usepackage{latexsym}
\newcommand{\be}{\begin{equation}}
\newcommand{\ee}{\end{equation}}
\newcommand{\bea}{\begin{eqnarray*}}

\newcommand{\rap}{\!\!\!&+&\!\!\!}
\newcommand{\ram}{\!\!\!&-&\!\!\!}

\newcommand{\eea}{\end{eqnarray*}}
\newcommand{\pa}{\partial}

\newcommand{\la}{\langle}
\newcommand{\ra}{\rangle}

\def\a{\alpha}
\def\b{\beta}

\def\d{\delta}

\def\f{\phi}
\def\g{\gamma}

\def\j{\psi}

\def\s{\sigma}

\def\v{\varphi}

\def\0{\over } \def\1{\vec } \def\2{{1\over2}} \def\4{{1\over4}}
\def\5{\bar } 
\def\6{\partial }
\def\7#1{{#1}\llap{/}}
\def\8#1{{\textstyle{#1}}} \def\9#1{{\bf{#1}}}
\def\.{\dot }
\def\^#1{\widehat{#1}}

\let\a=\alpha \let\b=\beta \let\g=\gamma \let\d=\delta

    \let\s=\sigma

  \let\D=\Delta

\begin{document}
\title{The Gross-Pitaevskii Equation for Bose
Particles in a Double Well Potential: Two Mode Models and Beyond}
\author{D. Ananikian}
\author{T. Bergeman}
\affiliation{Department of Physics and Astronomy, SUNY, Stony
Brook, NY 11794-3800}
\date{\today }
\begin{abstract}
There have been many discussions of two-mode models for Bose
condensates in a double well potential, but few cases in which
parameters for these models have been calculated for realistic
situations.  Recent experiments lead us to use the
Gross-Pitaevskii equation to obtain optimum two-mode parameters.
We find that by using the lowest symmetric and antisymmetric
wavefunctions, it is possible to derive equations for a more exact
two-mode model that provides for a variable tunneling rate
depending on the instantaneous values of the number of atoms and
phase differences. Especially for larger values of the nonlinear
interaction term and larger barrier heights, results from this
model produce better agreement with numerical solutions of the
time-dependent Gross-Pitaevskii equation in 1D and 3D, as compared
with previous models with constant tunneling, and better agreement
with experimental results for the tunneling oscillation frequency
[Albiez et al., cond-mat/0411757]. We also show how this approach
can be used to obtain modified equations for a second quantized
version of the Bose double well problem.
\end{abstract}
 \maketitle

\section{Introduction}

The analogy between double Bose condensates, separated by a barrier,
and Josephson oscillations of superconductors \cite{Josephson} was
apparently first suggested by Javanainen \cite{Juha1}, and has been
developed more thoroughly in a number of theoretical studies
\cite{Walls,Milburn,Smerzi,Zapata,Steel,Spekkens,Ivanov,Raghavan,Kivshar,Fran,Paraoanu,Leggett,Anglin,Mahmud1,Mahmud2,Links}.
Especially from work in \cite{Milburn,Smerzi,Raghavan} and more recently
in \cite{Fran,Mahmud2}, a rather elaborate
picture of phase space dynamics has now been developed.
The equations for tunneling oscillations of Bose
condensates in a double well potential have been
shown to resemble a pendulum whose length depends on the momentum.  In the
limit of small amplitude oscillations, the equations are the same as for
Josephson oscillations for superconductors separated by a weak
link \cite{Averin}.
It has also been shown that when atom-atom interactions exceed a critical value,
the ensemble will remain trapped in one well while the phase continually
increases, resembling a pendulum with sufficient energy to rotate.

Experiments showing interference when condensates in a potential with a
barrier were released \cite{Andrews} first stimulated interest in the problem
of Bose condensates in a double well potential.  More pertinent to the present
discussion are experiments that probe the evolution of the distribution
between two or more wells of an optical lattice.
Josephson oscillations have been observed in 1D optical potential arrays
\cite{Inguscio}. Recently for a double well potential, both the regimes of
tunneling and self-trapping of Rb atoms were observed \cite{Albiez}.  In
view of proposed extensions of these and other experimental
techniques, \cite{Peil,Zimmerman,Williams,Hu,Reichel}, it seems appropriate
now to reexamine the theory with the goal of developing models to deal with
more diverse conditions.

It is often assumed that the ``tight-binding'' approximation is valid,
leading to what is known as the Bose-Hubbard model \cite{Fisher,Jaksch},
or discrete nonlinear Schr\"{o}dinger equation \cite{Trombettoni,Rey}.
This model, which has been confirmed under the experimental conditions of
\cite{Inguscio}, employs parameters for tunneling and on-site energy
that are usually taken to be constant. One expects that with
sufficiently large numbers of atoms, the atom-atom repulsion will cause
the wavefunctions in a well to vary in size depending on the atom number,
and consequently, the tunneling parameter and onsite energy might vary.

We have found that it is possible to solve a more exact two-mode model
based on symmetric and antisymmetric solutions of the Gross-Pitaevskii
equation \cite{Kivshar,BAPS}. For weak interactions,
this new two-mode model produces negligible differences from previous
two-mode models. However, for larger interactions, there are substantial
differences and it turns out that the recent experiments \cite{Albiez}
begin to sample the
regime in which the differences are significant.  In this report, we show
that the new two-mode model implies a tunneling parameter that can vary
with time, depending on the number and phase of the ensemble in each well,
hence the name ``variable tunneling model'' (VTM).
Despite the additional terms needed to produce this result, the equations
eventually reduce to equations with the same form as the usual Bose-Josephson
junction equations, but with parameters defined differently, and with one
new term that can be significant in the case of strong interactions.
Below, we compare results obtained with this model to those with a
two-mode model with constant tunneling, with results of a multi-mode model, and
then with numerical solutions of the time-dependent Gross-Pitaevskii
equation (TDGPE). The parameters used in the two- or multi-mode models
are obtained from numerical solution of the stationary Gross-Pitaevskii
(GP) equation,
so it is perhaps not surprising that the model that mimics the GP equation
most closely also best reproduces results from the TDGPE.
For very large interactions, results from any two-mode
model will deviate from the TDGPE results, but agreement is most persistent
with the VTM. Under conditions of a particular experiment,
effects of non-zero temperature and experimental uncertainties may be
larger than the differences shown below.

The present study is primarily limited to a mean-field approach using the
GP equation, assuming that fluctuations and thermal excitations
are negligible and without quantizing particle number. Because atom number
is not quantized, the particle number difference and phase difference of atoms
in two wells or two modes are classical quantities in this approach.
Considerable theoretical effort has been devoted to the
second quantized form \cite{Milburn,Steel,Spekkens,Ivanov,Links,Mahmud2}.
We show below that our approach can be used to obtain more exact equations
for quantization, and we apply these equations to the case of weakly
interacting systems to show the connection between first and second quantized
theories in this limited regime. As shown in \cite{Mahmud2}, the classical
patterns appear clearly in the quantum phase space picture with as few
as 10 atoms. For experiments that involve on the order of 1,000 atoms,
it seems useful to have an improved GP (mean-field) approach.

An outline of this paper is as follows. Section II is devoted to 1D models.
We first (Section IIA) derive the
new (VTM) two-mode equations, and then (Section IIB) compare with previous
approaches.  Section IIC lays out a multi-mode approach, Section IID
discusses dynamics in phase space, Section IIE gives equations for a
second-quantized version.  Experiments are of course in 3D, with some
degree of transverse confinement.  Therefore in Section III we present
a formalism for 3D calculations and give a few results, including
comparisons with the experimental results of \cite{Albiez}.

\section{Models}  \label{Models1D}
\subsection{Time-Dependent Gross-Pitaevskii Equation}

When the temperature is sufficiently low and when particle numbers are
sufficient that second quantization effects are not important, the
time-dependent Gross-Pitaevskii
Equation (TDGPE) may be used for the wave function $\psi(x,t)$ for
interacting Bose condensate atoms at zero temperature in an external
potential $V_{ext}(x)$. Letting $\hbar = m = 1$,
a dimensionless version is
\begin{eqnarray} \label{tdgpe}
i \frac{\partial \psi}{\partial t} = - \frac{1}{2}
\frac{\partial^{2} \psi}{\partial x^{2}} + V_{\rm{ext}} \psi + g
|\psi|^{2}\psi.
\end{eqnarray}
The relationship between $g$ and $g_{3D}$ will be discussed in
section IIIA.   Here $\int dx|\psi(x,t)|^{2} = N$, where $N$ is
the number of atoms.  Except in Sec. \ref{secQ}, $N$ is not
quantized, and the approach is strictly mean-field.

We consider double-well potentials $V_{ext}(x)$ that are
symmetric in $x$.  Initially, we discuss one-dimensional versions.
Under the above conditions, we will
use results obtained with the TDGPE to test two-mode and multimode models
discussed below.  The TDGPE can tell us, for example, whether the phase
is nearly constant as a function of $x$ over an individual well.

\subsection{New Two Mode Model}

In many situations, a good approximation is obtained with a two mode
representation of $\psi(x,t)$. In early work \cite{Milburn},
wavefunctions localized in each well were used.  Later, $\pm$ combinations
of symmetric and antisymmetric functions, as in
\cite{Smerzi,Raghavan,Kivshar} provided a more accurate formulation,
and we follow the approach of \cite{Kivshar} here:
\begin{eqnarray}\label{appr}
\psi(x,t) &=& \sqrt{N}[\psi_{1}(t) \Phi_{1}(x) + \psi_{2}(t)
\Phi_{2}(x)];\label{anz}
\end{eqnarray}
\begin{eqnarray} \label{Phi12}
\Phi_{1,2}(x)& =& \frac{\Phi_{+}(x)
\pm \Phi_{-}(x)}{\sqrt{2}},
\end{eqnarray}
where
\begin{eqnarray} \label{Pnorm}
\Phi_{\pm}(x) = \pm \Phi_{\pm}(-x); \ \ \int dx \Phi_{i}\Phi_{j} =\delta_{i,j};
\ \ \ i,j=+,-.
\end{eqnarray}
The $\Phi_{\pm}$ will be assumed to be real, and to satisfy the
stationary GP equations
\begin{eqnarray} \label{modes}
\beta_{\pm} \Phi_{\pm} = - \frac{1}{2}  \frac{d^{2} \Phi_{\pm}}{d x^{2}} +
V_{ext} \Phi_{\pm} + \bar{g} |\Phi_{\pm}|^{2} \Phi_{\pm},
\end{eqnarray}
with $\bar{g}=gN$.

We can now define
\begin{eqnarray} \label{zphi}
z(t) &\equiv& |\psi_{1}(t)|^{2} - |\psi_{2}(t)|^{2},\\
\phi (t) &\equiv& \theta_{2}(t) - \theta_{1}(t).
\end{eqnarray}
here $\theta_{i}(t)$ are the phase arguments of the complex valued
function $\psi_{i}(t)$: $\psi_{i}(t)=
|\psi_{i}(t)|e^{i\theta_{i}(t)}$. The above normalization conventions
lead to a constraint on the $\psi_{i}(t)$:
\begin{eqnarray}
\int_{-\infty}^{\infty}dx|\psi|^{2} =N \Rightarrow |\psi_1(t)|^{2}
+ |\psi_2(t)|^{2} = 1.
\end{eqnarray}
Note that $\Phi_{1}(\Phi_{2})$ primarily occupies
the left(right) well, but has nonzero density on the other side.
In order to compare with results of the TDGPE, we define the
the number of atoms in the left well as follows:
\begin{eqnarray} \label{NL}
N_{L}\! = \! \int_{-\infty}^{0} \!\! dx|\psi(x,t)|^{2} = \frac{N}{2} +
Nz S,
\end{eqnarray}
and we define $S$ and $\Delta n$ as
\begin{eqnarray} \label{nt}
S \! = \!\! \int_{-\infty}^{0} \! dx \Phi_{+}(x) |\Phi_{-}(x)|; \ \ \
\Delta n = \frac{N_{L} - N_{R}}{N} = 2 z S.
\end{eqnarray}

From the ansatz~(\ref{anz}), the TDGPE (\ref{tdgpe}), and the GP
equation~(\ref{modes}), eventually one obtains differential equations
for $z$ and $\phi$.  We now briefly outline this derivation.
In the following, these quantities will be used:
\begin{eqnarray} \label{ABC}
\gamma_{ij} &=& \bar{g}\int \Phi_{i}^2(x) \Phi_{j}^2(x) \ dx; (i,j = +,-)
\nonumber \\
\Delta \gamma &=&  \gamma_{--} - \gamma_{++} \nonumber \\
\Delta \beta &=& \beta_{-} - \beta_{+} \nonumber \\
A &=& \frac{10 \gamma_{+-} - \gamma_{++} -\gamma_{--}}{4}  \nonumber \\
B &=& \b_--\b_++\frac{\g_{++}-\g_{--}}{2} = \Delta \beta -
\frac{\Delta \gamma}{2} \nonumber \\
C &=& \frac{\gamma_{++} + \gamma_{--} - 2 \gamma_{+-}}{4} \nonumber \\
F &=& \frac{\beta_{+} + \beta_{-}}{2}- \gamma_{+-}.
\end{eqnarray}
Substitution of (\ref{anz}) and (\ref{modes}) into (\ref{tdgpe}) yields
\begin{eqnarray}  \label{subb2}
i \frac{d \psi_{1}(t)}{d t} (\Phi_{+} + \Phi_{-})
+i \frac{d \psi_{2}(t)}{d t} (\Phi_{+} - \Phi_{-}) \hspace*{1.2cm} \nonumber \\
= \sum_{\pm} \left[ (\psi_{1}(t) \pm \psi_{2}(t)\right]
 \left[ \beta_{\pm} - gN|\Phi_{\pm}|^{2} \right] \Phi_{\pm}
\nonumber \\
+\frac{gN}{2} \sum_{\pm} \left( \Phi_{\pm}^{3} P_{\pm} + \Phi_{\pm}^{2}
\Phi_{\mp} Q_{\pm} \right),
\end{eqnarray}
where
\begin{eqnarray}
P_{\pm} \! &=& \!
 2(\psi_{1} \pm \psi_{2}) \! - \! |\psi_{1}|^{2} \psi_{1} \mp |\psi_{2}|^{2}\psi_{2}
\pm \psi_{1}^{2}\psi_{2}^{*} \! + \! \psi_{2}^{2}\psi_{1}^{*} \nonumber \\
Q_{\pm} \!&=& \!
\pm2(\psi_{2} \!-\! \psi_{1})\! +\! 5\psi_{1}|\psi_{1}|^{2}
\mp 5 \psi_{2}|\psi_{2}|^{2}
\pm \psi_{1}^{2} \psi_{2}^{*} \!-\! \psi_{2}^{2}\psi_{1}^{*}. \nonumber
\end{eqnarray}
The usefulness of the $\Phi_{\pm}$ basis is evident here, since integrals
with odd powers of $\Phi_{+}$ or $\Phi_{-}$ vanish. From the above equations,
including \ref{Pnorm}, the following equations for $\dot{\psi}_{1,2}(t)$
are obtained ($\mu,\nu=1,2;
\ \mu \neq \nu$):
\begin{eqnarray} \label{psi12}
i\frac{d\psi_{\mu}}{dt}&=& \left(F + A|\psi_{\mu}|^{2} -
\frac{\Delta \gamma}{4} \psi_{\mu}\psi_{\nu}^{*}\right) \psi_{\mu}
\nonumber \\
&+& \left(-\frac{\Delta \beta}{2} + \frac{\Delta \gamma}{4} |\psi_{\nu}|^{2} +
C \psi_{\mu}^{*} \psi_{\nu}\right) \psi_{\nu}
\end{eqnarray}

In analogous coupled equations presented elsewhere, as
in \cite{Milburn,Smerzi,Raghavan}, the coefficient of $\psi_{\nu}$ in the
equation for $\dot{\psi}_{\mu}$ is identified as the tunneling parameter,
and it is usually constant with time and independent of $\mu$. In the
above equation for $\dot{\psi}_{\mu}$, there are
additional terms in the coefficient of $\psi_{\nu}$ that are
functions of $\psi_1(t)$ and $\psi_{2}(t)$, hence varying with
time. Since $|\psi_{1,2}|^{2} = (1 \pm z)/2$ and
$\psi_{1}^{*}\psi_{2} = (1/2) \sqrt{(1-z^{2})}e^{i\phi}$, these
extra terms depend on instantaneous values of both $z$ and the phase
difference, $\phi$.
For this reason, we will call this model the ``variable
tunneling model,'' or VTM.  We will see that the additional
terms, although sometimes small, can bring this two-mode model
into closer agreement with solutions of the TDGPE.

Remarkably, despite some complexity of these additional terms,
relatively simple equations of familiar form can be obtained with no
approximations beyond the assumption of a two mode representation of
$\psi$, as in (\ref{anz}).  Equations (\ref{tdgpe}), (\ref{anz}),
(\ref{modes}), (\ref{subb2}) and (\ref{Phi12}) are used.  We obtain:
\begin{eqnarray}\label{zt}
\frac{d\phi}{dt} &=& A z + \frac{B z}{\sqrt{1 - z^{2}}} \cos{\phi}
- C z \cos{2 \phi}; \nonumber \\
\frac{dz}{dt} &=& -B \sqrt{1 - z^{2}} \sin{\phi} + C (1 -
z^{2})\sin{2 \phi}.
\end{eqnarray}
These equations have the same form as analogous equations in
\cite{Raghavan} except for the terms in $C$. They
can be written in Hamiltonian form
\begin{eqnarray} \label{dotzphi}
\dot{z}=-\frac{\partial H}{\partial \phi}, \ \ \ \
\dot{\phi}=\frac{\partial H}{\partial z}
\end{eqnarray}
with the Hamiltonian
\begin{eqnarray} \label{Ham}
H=A\frac{z^2}{2}-B\sqrt{1-z^2}\cos\phi+\frac12 C (1-z^2)\cos 2\phi
\end{eqnarray}
This Hamiltonian is an integral of motion for a classical system
with generalized coordinates ($z(t)$, $\f(t)$) and dynamical
properties~(\ref{zt}) and will be referred later as a classical
Hamiltonian. $H$ is not equal to the expectation value
$\frac{\la\j{\cal H}\j\ra}{\la\j\j\ra}$ of the quantum Hamiltonian
${\cal H}=-\2\frac{\pa^2}{\pa x} +V_{ext}(x)+g|\j|^2$ within two
mode approximation~(\ref{appr}). Since $\j(x,t)$ defined
as~(\ref{appr}) is not an eigenfunction of ${\cal H}$, the
expectation value $\frac{\la\j{\cal H}\j\ra}{\la\j\j\ra}$ is not
constant over time. However, the Hamiltonian~(\ref{Ham}) provides
information about dynamics in phase space, including
self-trapping, as will be discussed in section~\ref{PS}.

\begin{figure}
\includegraphics[scale=.52]{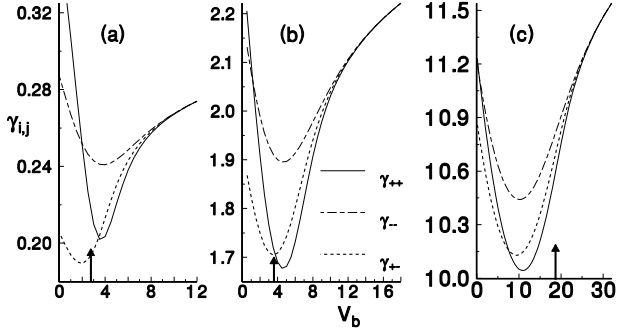}
\caption{Parameters $\g_{++},\g_{--},\g_{+-}$ as a function of $V_b$
for $gN$ = 1.0 (a); 10.0 (b); and 100.0 (c), calculated using (DVR) method
applied to the stationary GP equation~(\ref{modes}). The vertical
arrows denote values of $V_{b}$ for which $\beta_{+} = V_{b}$.}\label{gammas}
\end{figure}

In numerical work we often used a harmonic potential with Gaussian
barrier of varying height and width:
\begin{eqnarray}
V_{ext}(x)=\frac12 x^2+V_b e^{-(x/\sigma)^{2}}
\end{eqnarray}
Equations (\ref{anz}) and (\ref{modes}) imply that distances are scaled
by $\alpha_{x} = \sqrt{\hbar /M\omega_{x}}$, time by $1/\omega_{x}$
and energies, including $V_{b}$ above, by $\hbar \omega_{x}$, where
$M$ is the atomic mass, and $\omega_{x}/2\pi$ is the harmonic frequency.
This scaling will be used throughout this paper, and in particular in
all the figures.
To obtain numerical values for the overlap integrals $\gamma_{ij}$, where
$i,j = \pm$, we
solved (\ref{modes}) using the DVR method \cite{DVR,DVR1} with
increasingly finer mesh, with iterations for each mesh to make
the $\Phi_{\pm}$ functions and the the nonlinear term self-consisent.
Values for the
$\gamma_{ij}$ are shown as a function of barrier
height, $V_{b}$, for $\sigma$=1.5, for $gN$=1, 10 and 100, in Fig. \ref{gammas}.
For large enough $V_{b}$, all parameters $\gamma_{ij}$ are equal.
As $V_{b}$ decreases
from the asymptotic region, $\gamma_{++}$ decreases most rapidly because
$\Phi_{+}$, with no node, is less ``lumpy'' than $\Phi_{-}$.

\begin{figure}
\includegraphics[scale=.52]{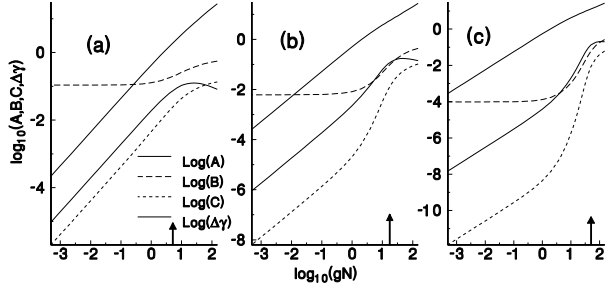}
\caption{$A, B, C$, and $\Delta \gamma$ parameters as a function
of $\bar{g} = gN$, on a log-log scale.  The three plots are for
$\sigma$ = 1.5, and $V_{b}$ = 4.0 (a); 7.0 (b); and 12.0 (c) in
units of $\hbar \omega_{z}$. The vertical arrows denote values of
$\log{\bar{g}}$ for which $\beta_{+} = V_{b}$.}\label{ABCfig}
\end{figure}

In Fig. \ref{ABCfig}, values for the parameters $A, B, C$ and $\Delta \gamma$
are shown.  For small $\bar{g}$, $B$ is nearly constant and close to the
value for a noninteracting gas, while $A$ and $C$ increase linearly with
$\bar{g}$.  For higher values of $V_{b}$, larger values of $\bar{g}$ lead
to more distortion of the $C$ and $\Delta \gamma$ parameters.
The parameter $C$ is several orders of magnitude smaller than
$A, B$ and $\Delta \gamma$ except when $\bar{g}$ is large compared to one.
When $C$ is much smaller than $\Delta \gamma$, it is justified to neglect
$C$ but preserve the difference between $B$ and $\Delta \beta$.

If the term in $C$ is neglected,
we have effectively derived an alternative Bose-Josephson junction (BJJ)
model, with revised parameters for tunneling phenomena between Bose
condensates in a double well potential.  It follows,
for example, that the discussion of Josephson plasmons in
\cite{Paraoanu} as Bogoliubov quasi-particles can be applied to the
present model, with appropriate substitutions.
However, when $\bar{g}$ is sufficiently large, the $C$ term cannot be
neglected.  We emphasize that this term comes strictly from the
nonlinear Gross-Pitaevskii equation for Bose condensates in a double well
potential and does not apply to superconducting Josephson junctions.

A useful estimate for the condition for a two-mode model to be valid is that
$\beta_{+} \leq V_{b}$.  With this rough criterion in mind, in many of the
plots, we will denote this point by vertical arrows. By this test, in
Fig. \ref{gammas}, two-mode models are valid to the right of the arrows, while
in Fig. \ref{ABCfig}, the regime of validity of two-mode models
is to the left of the arrows. In reality, the transition is not sharp, as
we will see below.

\subsection{Comparison with other model theories}

In \cite{Raghavan}, equations for $\dot{\psi}_{i}$ are derived with
help of orthogonal functions $\Phi_{1,2}(z)$, defined as in (\ref{modes})
above.  However, smaller terms were
neglected:  ``This approximation captures the dominant $z$ dependence
of the tunneling equations coming from the scale factors $\psi_{1,2} \approx
\sqrt{N_{1,2}}$, but ignores shape changes in the wavefunctions for
$N_{1}(t) \neq N_{2}(t)$.''  Effects due to shape changes were estimated to
be small. We find this to be the case in certain regimes but not always.

To make comparisons with the VTM, we write the
equations from \cite{Raghavan} taking $\hbar = m = 1$ as above:
\begin{eqnarray} \label{p1p2}
i \frac{d\psi_{1}}{dt} = (E_{1}^{0} + U_{1} |\psi_{1}|^{2}
)\psi_{1} - {\cal K} \psi_{2}; \nonumber \\
i \frac{d \psi_{2}}{dt} = (E_{2}^{0} + U_{2} |\psi_{2}|^{2}
)\psi_{2} - {\cal K} \psi_{1}; \nonumber \\
|\psi_{1}(t)|^{2} + |\psi_{2}(t)|^{2} = 1,
\end{eqnarray}
where, for $i$ = 1 or 2,
\begin{eqnarray}
E_{i}^{0} &=& \int dx \left[\frac{1}{2}|\nabla \Phi_{i}|^{2}
+ |\Phi_{i}|^{2} V_{ext}\right]; \nonumber \\
U_{i} &=& gN \int dx |\Phi_{i}|^{4}; \nonumber \\
{\cal K} &=& - \int dx \left[\frac{1}{2} (\nabla \Phi_{1} \nabla \Phi_{2})
+ \Phi_{1} V_{ext} \Phi_{2} \right].
\end{eqnarray}
Since the tunneling term (${\cal K}$) is constant with time,
we will refer to this model as the ``constant tunneling model'' (CTM).
In the CTM model, the functions $\Phi_{1,2}$ are related to symmetric
and antisymmetric functions $\Phi_{\pm}$ as in (\ref{modes}). However the
eigenvalues/chemical potentials $\beta_{\pm}$ do not directly apply.
To determine values for $E_{i}^{0}$ in terms of $\beta_{\pm}$
and $\gamma_{\pm,\pm}$, we introduce, for $i = +,-$,
\begin{eqnarray}
\epsilon_{i} = \int dx \left[ - \frac{1}{2}
\Phi_{i}\frac{d^{2} \Phi_{i}}{dx^{2}}
+ \Phi_{i} V_{ext} \Phi_{i} \right]  \nonumber \\
 = \beta_{i} - \bar{g} \langle |\Phi_{i}|^{4} \rangle
= \beta_{i} - \gamma_{ii}.
\end{eqnarray}
For the quantities $E_{1,2}^{0}$ introduced above, we obtain
\begin{eqnarray}
E^{0}_{1} = E^{0}_{2} = E
= \frac{\epsilon_{+} + \epsilon_{-}}{2}.
\end{eqnarray}
Furthermore,
\begin{eqnarray}
U_{1} = U_{2} &=& U = \bar{g}\langle |\Phi_{1,2}|^{4}\rangle =
\frac{\bar{g}}{4} \langle
|\Phi_{+} \pm \Phi_{-}|^{4} \rangle \nonumber \\
&=& \frac{1}{4} [ \gamma_{++} + 6 \gamma_{+-} + \gamma_{--} ] = A + 2C.
\hspace*{5mm} \end{eqnarray}
In the symmetric/antisymmetric basis, the coupling term becomes
\begin{eqnarray}\label{calK}
{\cal K}= \frac{\epsilon_{-} -\epsilon_{+}}{2} = \frac{\Delta \beta - \Delta
\gamma}{2} = \frac{B}{2} - \frac{\Delta \gamma}{4}.
\end{eqnarray}
The Hamiltonian is
\begin{eqnarray}\label{HCTM}
H_{\rm CTM} = U \frac{z^{2}}{2} - 2 {\cal K} \sqrt{1-z^{2}} \cos \phi.
\end{eqnarray}

In part, the differences in the two approaches arise because the wavefunctions, $\Phi_{1,2}$, extend somewhat into the opposite well, as noted above.
Rather than comparing the individual parameters, comparisons between the
two two-mode models are better performed in terms of properties that
are independent of the model, and may be calculated also with the
time-dependent Gross-Pitaevskii equation.  One such property is the
well-known Josephson plasma oscillation
frequency \cite{Leggett}, which is taken to be the oscillation frequency in the
limit of small amplitudes of $z$ and $\phi$.  Another derived property
is the onset of self-trapping
at $\phi=0$, which is usually labeled $z_{c}$, the critical value of $z$.
We will discuss this in subsection \ref{PS}.

In the limit of small $z$ and $\phi$, the equations for $\dot{z}$ and
$\dot{\phi}$ become
\begin{eqnarray} \label{omegas}
{\rm CTM}: \frac{dz}{dt} = - 2 {\cal K} \phi; \ \ \
\frac{d\phi}{dt} =  (U + 2 {\cal K})z; \hspace*{5mm} \nonumber \\
 \Rightarrow \ \omega_{oC}^{2} =
2 {\cal K}(U + 2 {\cal K}); \nonumber \\
{\rm VTM}:  \frac{dz}{dt} = (2C-B) \phi; \ \ \
\frac{d\phi}{dt} =  (A + B - C)z; \nonumber \\
 \Rightarrow \omega_{oV}^{2} = (B - 2C)(A + B - C).
\end{eqnarray}

\begin{figure}
\includegraphics[scale=.50]{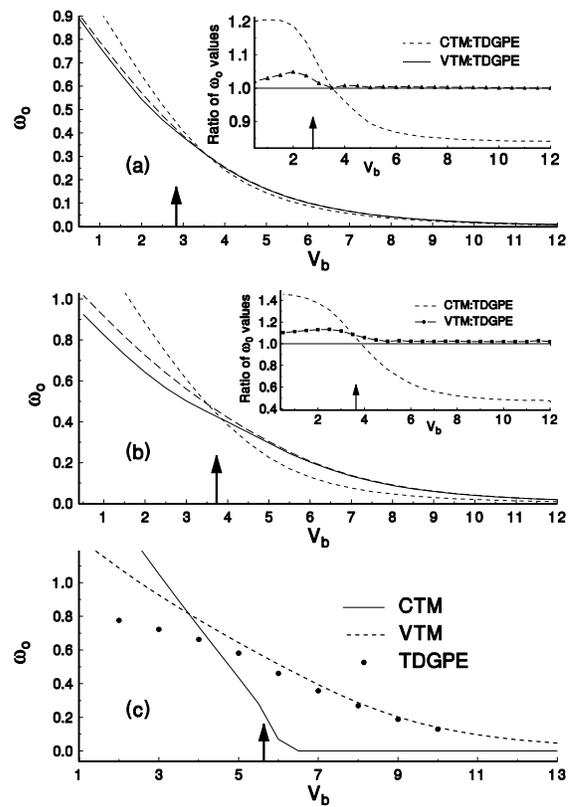}
\caption{\label{omegao}Comparisons of the oscillation frequencies for
small $z,\phi$ amplitude calculated from the CTM, the VTM and the TDGP
equation, for $gN$ = 1.0 (a); 3.0 (b); 10.0 (c). The insets in (a) and
(b) show ratios of CTM and VTM results to TDGPE results. Vertical arrows
denote values of $V_{B}$ for which $V_{b} = \beta_{+}$.}
\end{figure}

In every 1D case we have considered, $B - 2C >0$ and $A + B - C >0.$
Numerical results obtained with the VTM and CTM models are shown in
Fig. \ref{omegao} in comparison with frequencies obtained with the TDGP
equation.  For $\bar{g} \ll 1$, all three approaches agree well.
For $\bar{g}$=1 and large $V_{b}$, the
values for $\omega_{o}$ from the CTM are about 16\% less than from the VTM,
while for $\bar{g}$=3, the asymptotic difference is about a factor of two.
For larger values of $\bar{g}$ and for large $V_{b}$, as
illustrated for $\bar{g}$=10 in Fig. \ref{omegao}c, ${\cal K}$
becomes negative, hence $\omega_{oC}$ becomes imaginary, and the
real part of $\omega_{0}$ plotted in Fig. \ref{omegao} is zero.

\begin{figure}
\includegraphics[scale=.45]{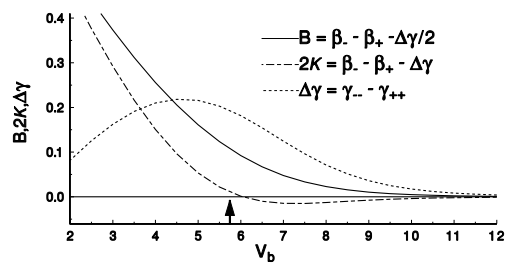}
\caption{\label{BK}Values for the parameters $B, 2{\cal K}$, and
$\Delta \gamma = \gamma_{--} - \gamma_{++}$ for $\bar{g}$ = 10 and
values of the barrier height, $V_{b}$ as indicated. Although ${\cal K}$
becomes negative for $V_{b} > 6$, $B$ remains positive.}
\end{figure}

Values of $B, 2{\cal K}$, and $\Delta \gamma$ for $\bar{g} = 10$ and
$\sigma_{z}$ = 1.5 are shown in Fig. \ref{BK} (2{\cal K} is plotted
because (\ref{calK}) shows
that in the limit $\gamma_{ij} \rightarrow 0$, $B = 2 {\cal K}$, and from
(\ref{omegas}), we see that $B$ plays a role equivalent to
$2 {\cal K}$). The region where
${\cal K} < 0 $ is clearly indicated.  $\beta_{\pm}$ are the actual
eigenvalues, which are calculated with the nonlinear interaction terms
included. The quantities $\epsilon_{\pm}$ have no direct physical meaning,
so it is not surprising that they can lead to anomalous results.
Note also that the putative regime of validity of two-mode models
is to the right of the vertical arrows in each figure, and that for
$\bar{g}$=10, ${\cal K}$ is negative over most of this region.


Thus from calculations of the Josephson plasma oscillation frequencies,
we conclude that the additional terms derived in the VTM
model take better account of nonlinear interaction effects and produce
better agreement with full TDGPE results.  For
low atomic numbers and weak interactions, these additional terms are
not needed. It is also evident that as interactions increase in magnitude,
neither two-mode model reproduces TDGPE results quantitatively. This will
lead us to examine multimode models below.

\begin{figure}
\includegraphics[draft=false, width=3.6in,keepaspectratio]{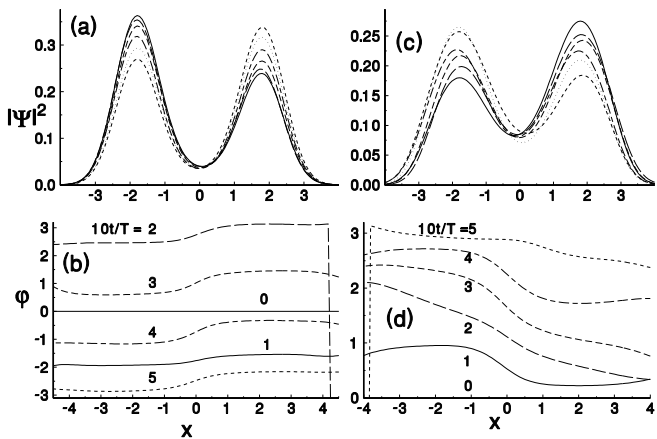}
\caption{(a) and (c) Evolution of $|\Psi(x)|^{2}$ over one-half cycle
of tunneling oscillation. (b) and (d) Evolution of phase, $\phi(x)$
under the same conditions as in (a) and (c), respectively.
The conditions for (a) and (b) are $\bar{g}$ = 2.0, $V_{b}$ = 5.0;
for (c) and (d), $\bar{g} = 10.0, V_{b} = 5.0$, and in each case
$\sigma$ = 1.5. In (a) and (c), the initial function $|\Psi(x)|^{2}$ is
denoted by thick solid lines.  In (b) and (d), the initial value of the
phase is everywhere zero. In (d) after the initial time, the phase is
clearly not uniform over either
well because of the strong interactions and low barrier.} \label{PP}
\end{figure}

First, however, it will be helpful to take another perspective by looking
at results simply from the TDGPE.  Fig. \ref{PP} shows $|\Psi(x)|^{2}$
and $\phi(x)$ as they evolve over one-half cycle under conditions in which
(in a and b) the phase is nearly constant over each well,
and (in c and d) with a larger $\bar{g}$ interaction parameter such that
the phase over each well is not constant at a given time.
In the latter case, the phase difference cannot be defined, and
any two-mode model fails.

\subsection{Multimode approximation}

From Fig. \ref{omegao}, we saw that there are deviations in the Josephson
plasma frequency,
$\omega_{0}$, between even the more exact (VTM) two-mode model and numerical
solutions of the TDGPE. These deviations raise the question whether better
agreement can be obtained by expanding the set of basis functions beyond
simply $\Phi_{+}$ and $\Phi_{-}$.

In this section we introduce a generalization of the VTM
two-mode model. Starting from the TDGP equation
\begin{eqnarray} \label{gpemul}
i \frac{\partial \psi}{\partial t} = - \frac{1}{2}
\frac{\partial^{2} \psi}{\partial x^{2}} + V_{\rm{ext}} \psi + g
|\psi|^{2}\psi
\end{eqnarray}
we introduce the following ansatz
\be\label{anzat}
\j(x,t)= \sqrt{N} \sum_{k=0}^{N-1}b_k(t)e^{-i\b_k t}\f_k(x)
\ee
where $\f_k(x)$ satisfy the following equations
\begin{eqnarray}\label{mulmod}
\beta_{2s} \phi_{2s} = - \frac{1}{2}  \frac{d^{2} \phi_{2s}}{d
x^{2}} + V_{ext} \phi_{2s} + \bar{g} |\phi_{0}|^{2} \phi_{2s}\\
\beta_{2s+1} \phi_{2s+1} = - \frac{1}{2}  \frac{d^{2}
\phi_{2s+1}}{d x^{2}} + V_{ext} \phi_{2s+1}\hspace{0.5cm} \\
+ \bar{g} |\phi_{1}|^{2} \phi_{2s+1}.\nonumber
\end{eqnarray}
Thus $\phi_{0,1} = \Phi_{\pm}$ as defined above,
with normalization $\int dx \f_{i}(x)\f_j(x)=\d_{ij}$. Here we are
effectively using the virtual excited states of the Gross-Pitaevskii
equation rather than Bogoliubov quasi-particle states.
Equilibrium thermodynamics is not the goal here. Any orthonormal basis
offers an extension of the two-mode model, and the quasi-particle
basis is unnecessarily cumbersome for this application. Substituting
the ansatz~(\ref{anzat}) into the GP equation~(\ref{gpemul}) and
using equations for $\f_{2s}(x)$ and $\f_{2s+1}(x)$ with the
orthogonality property, we obtain the following equation for the
time depending amplitudes $b_r(t)$.
\be\label{equa}
\begin{array}{rcl}
i\.b_r= \ram \sum\limits_j b_{2j}\g_{00,2j,r}e^{i(-\b_{2j}+\b_r)t} \\
\\
\ram\sum\limits_j
b_{2j+1}\g_{11,2j+1,r}e^{i(-\b_{2j+1}+\b_r)t}\\
\\
\rap
\sum\limits_{s,n,m}b_nb^*b_s\g_{nmsr}e^{i(-\b_n+\b_m-\b_s+\b_r)}
\end{array}
\ee
These are $2J$ equation for real functions $|b_j(t)|$ and
$arg(b_j(t))$, where $J$ is number of modes. However there is the
following constraint: $\sum\limits_j|b_j(t)|^2=1$ which is a
consequence of the normalization condition for the wave function
$\j(x,t)$. Since also the overall phase is arbitrary, we effectively
have $2J-2$ equations for $2J-2$ independent variables.  Therefore,
we define $b_j(t)=c_j(t)e^{i\a_j(t)}$, with $c_j(t)=|b_j(t)|$, and introduce
the following variables
\bea
\D_r=c_0^2-c_r^2, \quad r=1,\dots,J-1\\
\v_r=\a_{r-1}-\a_r, \quad r=1,\dots,J-1
\eea
It is not difficult to restate equations~(\ref{equa}) in terms of the new
variables.

As in the case of VTM two-mode model, the main ingredients of
multimode approximation are parameters $\g_{klmn}$ that can be
found numerically from eigenfunctions of the Gross-Pitaevskii operator for
the symmetric and antisymmetric ``condensates.''
In making comparisons with two-mode model results and with numerical
solutions of the TDGPE, we will use the number difference
$\Delta n(t)$ defined in (\ref{nt}), rather than $z(t)$, which is
not defined for the TDGPE.  As an initial condition for the TDGPE,
we use desired linear combinations of $\Phi_{\pm}$ (relabeled $\phi_{0,1}$
in (\ref{mulmod})).  In a given experimental situation, the actual initial
condition might differ and might need to be modeled more precisely.

\begin{figure}
\includegraphics[draft=false, width=3.5in,keepaspectratio]{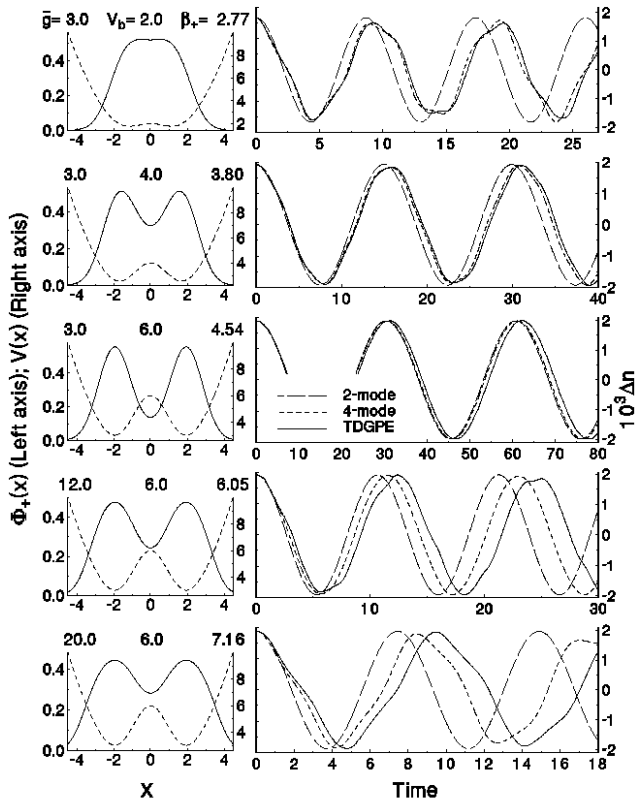}
\caption{Left column: The functions $\Phi_{+}(x)$ and potentials $V(x)$ for
conditions indicated above each frame: $\bar{g}, V_{b}$, with $\sigma_{z}$
= 1.5 in each case. The last figure gives the chemical potential, $\beta_{+}$.
Right column: time evolution of the fractional number difference,
$\Delta n$ times $10^{3}$ for very small initial imbalance.
The three curves are obtained from the 2-mode VTM model (long dashes),
the 4-mode model (short dashes) and the TDGPE (solid curve).} \label{timevol}
\end{figure}

What our results show generally is that in circumstances in which the VTM
differs significantly from TDGPE, the time evolution curve is not sinusoidal,
but is distorted by higher frequency components. Therefore one cannot easily
extract a single frequency, for example, to correct the discrepancies
exhibited in Fig. \ref{omegao}. Figure \ref{timevol} shows the
actual time evolution curve for several cases. These curves should be viewed
in light of the statement \cite{Rey} that when $\beta_{+} < V_{b}$, the
tight-binding limit applies, or in our case, the two-mode VTM applies. As
shown in this figure, the two-mode model agrees quite well with the TDGPE
curve for $\bar{g}=3.0, V_{b} = 6.0$, for which $\beta_{+} = 4.54$ is less
than $V_{b}$. For larger $\bar{g}$ or smaller $V_{b}$, the 2-mode and TDGPE
currves differ both in frequency and shape. In each of these cases, results
obtained with a 4-mode model yield better agreement with the TDGPE curves.
It is remarkable that this good agreement appears even for a very
low barrier, $V_{b} = 2.0$, for $\bar{g}$=3.0.

\subsection{Phase space dynamics} \label{PS}

The evolution of $z,\phi$ from the coupled equations, (\ref{zt}),
closely resembles the dynamical evolution phenomena thoroughly
discussed in \cite{Raghavan}. We give a brief review to point out the
differences arising from use of the VTM.

\begin{figure}
\includegraphics[draft=false, width=3.5in,keepaspectratio]{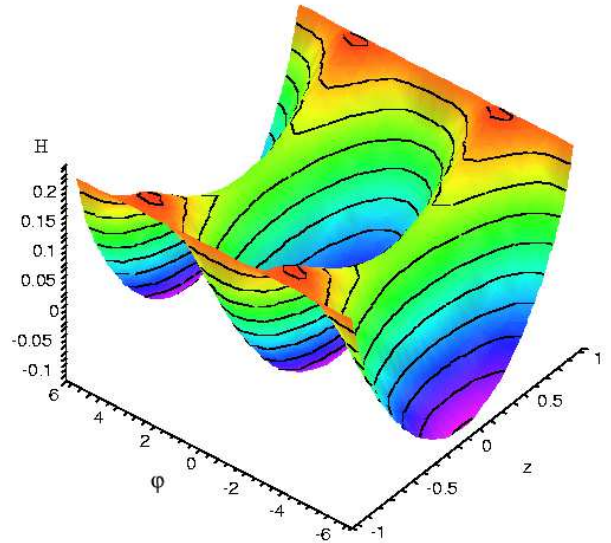}
\caption{Hamiltonian surface $H(z,\f)$ for $V_0=4$, $\s=1.5$,
$\bar{g}=1$. Trajectories lie on the surface, following contour lines
that represent constant energy levels.}\label{hamsur}
\end{figure}

To visualize the dynamics, it is helpful to view a plot of the Hamiltonian
surface, $H(z,\phi)$, as shown in Fig. \ref{hamsur} for generic
values of $A, B$ and $C$. The surface is periodic in $\phi$, with minima
at $z=0, \phi=2n \pi$ and saddle points or maxima at $z=0, \phi = (2n+1)\pi$,
where $n$ is an integer. Trajectories are horizontal curves (constant $H$)
lying on this surface.

Within either two-mode model, self trapping occurs for $H$ above
$H_s$, the value of classical Hamiltonian at the saddle point.
Critical values of $z = z_{c}$ are defined as values of
$z(\phi=2n\pi) = z_{0}$ that give $H(z,\f)$ equal to $H_{s}$. For
$|z_{0}|
> z_{c}$, trajectories will not pass through $z=0$ and $z$ will
remain positive or negative. For the VTM model, the Hamiltonian
given by (\ref{Ham}), gives
\begin{eqnarray}
H_{s} = H(0,\pi) = B + \frac{C}{2} = H(z_{c},0).
\end{eqnarray}
From this result and (\ref{Ham}), we obtain
\begin{eqnarray}  \label{zcv}
z_{c,V} =  \frac{2}{A-C}\left[ B(A-B-C) \right]^{1/2}.
\end{eqnarray}

For the CTM model, the Hamiltonian of (\ref{HCTM}) yields
\begin{eqnarray}
H_{s} = H(0,\pi) = 2{\cal K} = H(z_{c},0),
\end{eqnarray}
so that
\begin{eqnarray} \label{zcc}
z_{c,C} = \frac{2}{U}[2{\cal K}(U - 2 {\cal K})]^{1/2}.
\end{eqnarray}
Here the model breaks down when either ${\cal K} < 0$ (see Fig.~\ref{BK})
or $U - 2 {\cal K} < 0$.

\begin{figure}
\includegraphics[draft=false, width=3.6in,keepaspectratio]{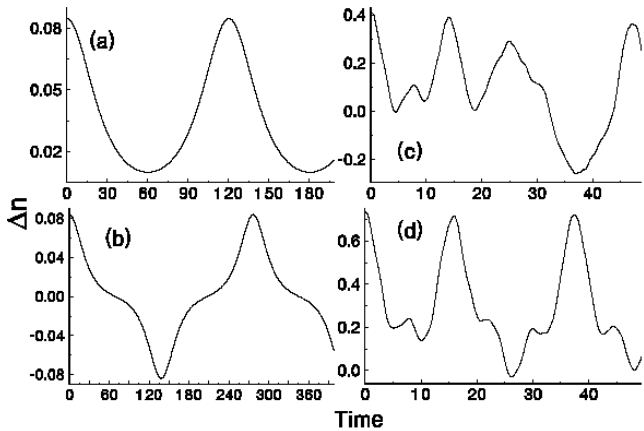}
\caption{Temporal evolution of $\Delta n$ for various cases: (a)
$\bar{g} =3.0, V_{b} = 9.0, z_{0} = \Delta n(0) =0.0846$; (b) $\bar{g} = 3.0,
V_{b} = 9.0, \Delta n(0) = 0.0838$; (c) $\bar{g} = 10.0, V_{b} = 5.5,
\Delta n(0) = 0.411$; (d) $\bar{g} = 3.0, V_{b} = 4.0, \Delta n(0) =
0.735$.} \label{nlnr}
\end{figure}

Before presenting results of calculations of $z_{c}$, we recognize
that as $|z| \approx |\Delta n|$ and $\bar{g}$ increase, as in
Fig. \ref{timevol}, higher modes enter. The variation of $\Delta
n$ with time becomes irregular rather than close to sinusoidal, as
shown by several plots obtained from calculations with the TDGPE
in Fig. \ref{nlnr}. For Fig. \ref{nlnr}a and \ref{nlnr}b
(differing very slightly in $\Delta n(0)$, but on opposite sides
of $\Delta n = z_{c}$), closely resemble results one would expect
from a two-mode model.  Figure \ref{nlnr}c and d show irregular
curves from the TDGPE in a regime where the two-mode model does
not apply.  In Fig. \ref{nlnr}c, there are oscillations of $\Delta
n$ within the range $\Delta n >0$ before $\Delta n$ eventually
becomes $< 0$. Fig. \ref{nlnr}d shows that $\Delta n <0$ is
achieved for only a brief duration (between T = 25 and 29).
Neither of these cases can be considered ``self-trapping,'' but
they are far removed from symmetric, periodic oscillations. Under
such conditions of low barrier and/or strong interactions, it is
somewhat arbitrary to make the distinction between self-trapping
and not self-trapping.

\begin{figure}
\includegraphics[draft=false, width=3.5in,keepaspectratio]{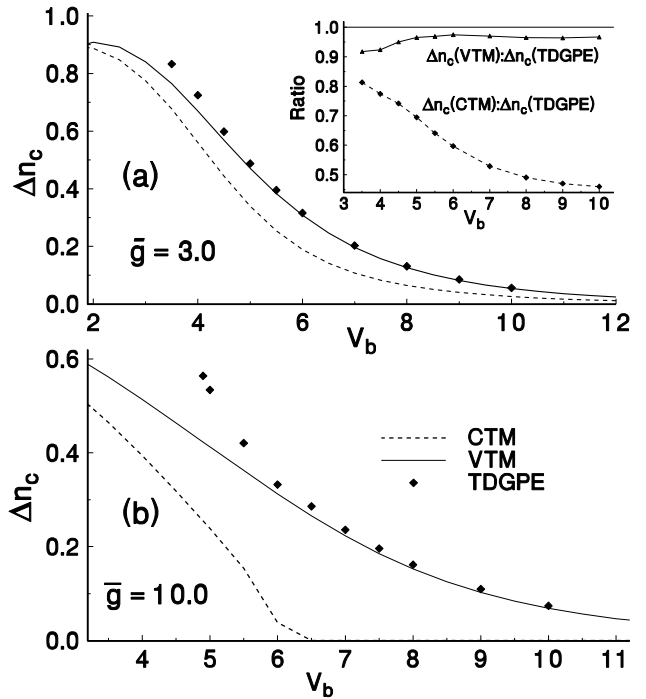}
\caption{Values for $\Delta n_{c}$ from the CTM, VTM and TDGPE models, for
a) $\bar{g}$ = 3.0 and b) $\bar{g}$ = 10.0.  The inset in (a) shows the
ratio of CTM and VTM values to TDGPE results.}\label{zcfig}
\end{figure}

Nonetheless, we have attempted to establish criteria and apply them
consistently so as to compare results from the CTM, VTM, and TDGPE
approaches, as shown in Fig. \ref{zcfig}. Here, $z_{c}$ values from
(\ref{zcv}) and (\ref{zcc}) have been restated in terms of $\Delta n_{c}$
using (\ref{nt}) in order to compare with TDGPE results.  For both
$\bar{g}$ = 3.0 and 10.0, when $V_{b}$ is high enough, there is good
agreement between VTM and TDGPE results.  CTM results are significantly
lower for $\bar{g}$=3.0, while for $\bar{g}$=10.0, as in
Fig. \ref{omegao}, the fact that ${\cal K}$ becomes negative invalidates
this approach in this regime of strong interactions.

In the self-trapping regime, maximal and minimal values of $z(t)$ can be
obtained by solving the equation $\dot{z}=0$:
\begin{eqnarray}
z=\pm\sqrt{1-\left[\frac{|B|}{C-A}\pm\sqrt{\left(\frac{B}{C-A}\right)^2-\frac{2H-A}{C-A}}\right]^2}.
\end{eqnarray}
Plus or minus signs in front of the square root correspond to
different initial conditions for $z$ (positive or negative
respectively).  An elegant discussion of dynamics, and separatrices, in
phase space is given in \cite{Fran} (explicitly for the case $C$=0).

In \cite{Raghavan}, it was pointed out that closed trajectories
on the surface of $H$ can also occur around maxima on the lines
$\phi = (2n+1)\pi$.
These are the so-called $\pi$-phase modes. For the VTM, the condition
for these maxima is that  $|B|<|(A+C)|$.  The actual values, $z_{\pi}$,
at which these maxima occur can vary drastically from one model to
the other.

Even for the case of negligibly small overlap, the momentum $z(t)$ in
the VTM differs drastically from the CTM when conditions place these
two models on opposite sides of the transition to self-trapping.
Far from the neighbourhood of the self trapping transition in
$\bar{g},\sigma,V_0$, the differences are less.

\subsection{Second Quantization}\label{secQ}

Previous discussions of quantized versions of the Bose double well
problem \cite{Milburn,Ivanov} were valid to first order in the overlap
of the wavefunctions in each well. Also in the recent work \cite{Mahmud2},
certain approximations are made for the wavefunction overlap. Now that we
have a mechanism for treating the overlap more exactly,
there are new possibilities for extending the regime of
validity of quantum approaches, which are necessarily based on two-mode
models.

The energy functional describing trapped BEC in terms of creation
and annihilation operators $\psi(x,t)$, $\psi^\dag(x,t)$ can be written
\begin{eqnarray}\label{Hq}
H_{2}=H_{0} + H_{1}; \ \ \
H_{0} &=& \int dx \left[ -\frac{1}{2} \hat{\psi}^{\dagger}
\nabla^{2} \hat{\psi} +\hat{\psi}^{\dagger} V_{ext}  \hat{\psi} \right];
\nonumber \\
H_{1} &=&\frac{g}{2} \int dx \hat{\psi}^{\dagger} \hat{\psi}^{\dagger}
 \hat{\psi} \hat{\psi},
\end{eqnarray}
with the commutator
$[\hat{\psi}(x,t),\hat{\psi}^\dag(x',t)]=\delta(x-x')$.

As above, we will characterize the time evolution in terms of two modes
that are predominantly (but not exclusively) located in the left and right
wells.  However, the derivation is easier when written in terms
of the symmetric and antisymmetric functions, $\Phi_{\pm}$ rather than
in terms of $\Phi_{1,2}$, because $\langle \Phi_{\pm}^{3}\Phi_{\mp} \rangle
=0$, whereas $\langle \Phi_{1,2}^{3} \Phi_{2,1} \rangle \neq 0$.
We therefore write a ``mixed basis'' expression:
\begin{eqnarray} \label{psiq}
\hat{\psi} = \frac{1}{\sqrt{2}}
\left[ \hat{c}_{1} (\Phi_{+} + \Phi_{-})
+ \hat{c}_{2} (\Phi_{+} -  \Phi_{-}) \right],
\end{eqnarray}
in which
\begin{eqnarray}
\hat{c}_{1,2} = \frac{1}{\sqrt{2}} \int dx \hat{\psi}(\Phi_{+} \pm \Phi_{-})
\end{eqnarray}
are projections of $\hat{\psi}$.  $\Phi_{\pm}$ are solutions to the GP
equation as above. In particular, the following form will be useful:
\begin{eqnarray} \label{Phipm}
-\frac{1}{2} \nabla^{2} \Phi_{i} + V_{ext} \Phi_{i} = \beta_{i} \Phi_{i} - gN
|\Phi_{i}|^{2}|\Phi_{i}.
\end{eqnarray}
Also we have $[\hat{c}_{i},\hat{c}^{\dagger}_{j}] = \delta_{ij}.$

Substituting \ref{psiq} and \ref{Phipm} into the above equation for
$H_{0}$, we obtain four terms:
\begin{eqnarray}\label{ham1}
H_{0}= \frac{1}{2} \left\{\left[\hat{c}_{1}^{\dagger} \hat{c}_{1} +
\hat{c}_{2}^{\dagger} \hat{c}_{2}\right] \left[
\beta_{+} - \gamma_{++} + \beta_{-} -  \gamma_{--}\right] \right. \nonumber \\
\left. + \left[\hat{c}_{1}^{\dagger} \hat{c}_{2} +
\hat{c}_{2}^{\dagger} \hat{c}_{1}\right] \left[
\beta_{+} - \gamma_{++} - \beta_{-} +  \gamma_{--}\right] \right\}
\end{eqnarray}
Upon substituting \ref{psiq} into the above equation for $H_{1}$, we
obtain 16 terms, each with products of two creation and two annihilation
operators, times integrals of the form
\begin{eqnarray}
(+)^{i}(-)^{j} = \frac{g}{2} \int dx (\Phi_{+} + \Phi_{-})^{i}
(\Phi_{+} - \Phi_{-})^{j}.
\end{eqnarray}
In particular
\begin{eqnarray} \label{H1}
H_{1} =\hat{c}_{1}^{\dagger} \hat{c}_{1}^{\dagger} \hat{c}_{1}
\hat{c}_{1} (+)^4
+ \hat{c}_{2}^{\dagger} \hat{c}_{2}^{\dagger} \hat{c}_{2} \hat{c}_{2}
(-)^{4} \hspace*{2cm} \\
+ \left[
 \hat{c}_{1}^{\dagger} \hat{c}_{1}^{\dagger} \hat{c}_{1} \hat{c}_{2}
+\hat{c}_{1}^{\dagger} \hat{c}_{1}^{\dagger} \hat{c}_{2} \hat{c}_{1}
+\hat{c}_{1}^{\dagger} \hat{c}_{2}^{\dagger} \hat{c}_{1} \hat{c}_{1}
+\hat{c}_{2}^{\dagger} \hat{c}_{1}^{\dagger} \hat{c}_{1} \hat{c}_{1} \right]
(+)^3(-) \nonumber \\ + \left[
\hat{c}_{2}^{\dagger} \hat{c}_{1}^{\dagger} \hat{c}_{1} \hat{c}_{2}
+\hat{c}_{2}^{\dagger} \hat{c}_{1}^{\dagger} \hat{c}_{2} \hat{c}_{1}
+\hat{c}_{2}^{\dagger} \hat{c}_{2}^{\dagger} \hat{c}_{1} \hat{c}_{1}
\right. \hspace*{2cm} \nonumber \\ \left.
+\hat{c}_{1}^{\dagger} \hat{c}_{2}^{\dagger} \hat{c}_{1} \hat{c}_{2}
+\hat{c}_{1}^{\dagger} \hat{c}_{2}^{\dagger} \hat{c}_{2} \hat{c}_{1}
+\hat{c}_{1}^{\dagger} \hat{c}_{1}^{\dagger} \hat{c}_{2} \hat{c}_{2}
\right](+)^{2}(-)^2 \nonumber \\ +\left[
\hat{c}_{2}^{\dagger} \hat{c}_{2}^{\dagger} \hat{c}_{1} \hat{c}_{2}
+\hat{c}_{2}^{\dagger} \hat{c}_{2}^{\dagger} \hat{c}_{2} \hat{c}_{1}
+\hat{c}_{2}^{\dagger} \hat{c}_{1}^{\dagger} \hat{c}_{2} \hat{c}_{2}
+\hat{c}_{1}^{\dagger} \hat{c}_{2}^{\dagger} \hat{c}_{2} \hat{c}_{2}
\right](+)(-)^3 \nonumber \\
= \sum_{i=0}^{4} D(i) (+)^{4-i} (-)^{i} \hspace*{2cm}
\end{eqnarray}
Recalling the definitions
\begin{eqnarray}
4U = \gamma_{++} + 6 \gamma_{+-} + \gamma_{--}; \ \ \
4C = \gamma_{++} + \gamma_{--} - 2\gamma_{+-}; \nonumber \\  A = U - 2C; \ \ \
\Delta \gamma = \gamma_{--} - \gamma_{++}; \ \ \ \Delta \beta = \beta_{--}
- \beta_{++},
\end{eqnarray}
we obtain
\begin{eqnarray}
(\pm)^4 & = &\frac{g}{2} \int dx (\Phi_{+}\pm\Phi_{-})^{4}
\nonumber \\
&=& \frac{1}{2N}(\gamma_{++} + 6 \gamma_{+-} + \gamma_{--}) = \frac{2U}{N}; \\
(\pm)^3(\mp) &=& \frac{g}{2} \int dx (\Phi_{+}^2- \Phi_{-}^{2})(
\Phi_{+}^{2} \pm 2\Phi_{+} \Phi_{-} + \Phi_{-}^{2})  \nonumber \\
&=&\frac{1}{2N} (\gamma_{++} - \gamma_{--}) = -\frac{1}{2N} \Delta \gamma.
\nonumber \\
(+)^2(-)^2 &= &\frac{g}{2} \int dx (\Phi_{+}^{2} - \Phi_{-}^{2})^{2}
\nonumber \\
&=& \frac{1}{2N}(\gamma_{++} - 2 \gamma_{+-} + \gamma_{--})= \frac{2C}{N}.
\end{eqnarray}

We wish to represent $H_{2}$ in terms of the following operators:
\begin{eqnarray}
N = N_{1} + N_{2} = \hat{c}_{1}^{\dagger}\hat{c}_{1}
+ \hat{c}_{2}^{\dagger}\hat{c}_{2}; \ \ \
J_{x} = \frac{1}{2}(c_{2}^{\dagger} \hat{c}_{2} - c_{1}^{\dagger} \hat{c}_{1})
\nonumber \\
J_{y} = \frac{i}{2}(c_{2}^{\dagger} \hat{c}_{1} - c_{1}^{\dagger} \hat{c}_{2})
; \ \ \
J_{z} = \frac{1}{2}(c_{2}^{\dagger} \hat{c}_{1} + c_{1}^{\dagger} \hat{c}_{2}),
\end{eqnarray}
and Casimir element $J^2=\frac{\hat N}{2}\left(\frac{\hat N}{2}+1
\right)$, so that
\begin{eqnarray}
[J_i,J_j]=i\epsilon_{ijk}J_k.
\end{eqnarray}
Also we will need
\begin{eqnarray}
 N_{1} N_{2} = \frac{N^{2}}{4} - J_{x}^{2}.
\end{eqnarray}
Then the products of four annihilation and creation operators, the
$D(i)$ terms in (\ref{H1}), reduce to:
\begin{eqnarray}\label{ham2}
D(0)+D(4) = (\hat{c}_{1}^{\dagger})^{2} \hat{c}_{1}^2 +
(\hat{c}_{2}^{\dagger})^{2} \hat{c}_{2}^2 = \frac{N^{2}}{2} - N + 2 J_{x}^{2};
\nonumber \\
D(1) + D(3) = 4 (N-1) J_{z}; \hspace*{2cm} \nonumber \\
D(2) = 4 J_{z}^{2} + 2 N_{1} N_{2} = 4 J_{z}^{2} + \frac{N^{2}}{2}
- 2 J_{x}^{2}.
\end{eqnarray}
Collecting terms, neglecting terms that are constant, we obtain
\begin{eqnarray}
H_{2} = -J_{z} \left(\Delta \beta + \Delta \gamma - \frac{2\Delta \gamma}{N}
\right) + \frac{4(A+C)}{N} J_{x}^{2} + \frac{8C}{N} J_{z}^{2}.
\end{eqnarray}

The quantum equations of motion read
\begin{eqnarray}
\dot{J_i}=i[\hat{H},J_i],
\end{eqnarray}
which yields
\begin{eqnarray}
\dot{J_x}&=&\left(\Delta \beta +\Delta \gamma -\frac{2\Delta \gamma}{N}
\right)J_y- \frac{8C}{N} (J_yJ_z+J_zJ_y) \nonumber \\
\dot{J_y}&=&-\left(\Delta \beta +\Delta \gamma -\frac{2\Delta \gamma}{N}
\right)J_x  \nonumber \\
&& - \frac{8C}{N} (J_xJ_z+J_zJ_x)-\frac{4(A+C)}{N}(J_zJ_x+J_xJ_z)
\nonumber \\
\dot{J_z}&=& \frac{4(A+C)}{N}(J_yJ_z+J_zJ_y)
\end{eqnarray}

The above Hamiltonian, $H_{2}$, is to be compared with expressions derived
previously \cite{Milburn,Ivanov,Leggett,Links,Fran,Mahmud2}. Although a 
general second quantized Hamiltonian was written many years ago by nuclear 
physicists \cite{LKM} (since known as the LMG model), most applications
seem to involve simply the terms in $J_{z}$ and $J_{x}^2$.
Using the operators defined above and assuming a symmetric
double well potential, the expression in \cite{Leggett}, for example, can be
written:
\begin{eqnarray}
H_{\rm canon} = - {\cal E}_{J} J_{z} + \frac{1}{2}K J_{x}^{2}.
\end{eqnarray}
The comparison provides the following translation:
\begin{eqnarray}
{\cal E}_{J} = \Delta \beta + \Delta \gamma - \frac{2 \Delta \gamma}{N};
\ \ \ K = \frac{8(A+C)}{N}.
\end{eqnarray}
The regimes defined in \cite{Leggett} then become (neglecting the
$2 \Delta \gamma/N$ term:
\begin{eqnarray}
{\rm Rabi}&:& \ \frac{K}{{\cal E}_{J}} \ll \frac{1}{N} \Rightarrow
R \ll 1; \ \ \ R = \frac{8(A+C)}{\Delta \beta + \Delta \gamma}; \nonumber \\
{\rm Josephson}&:& \ \ \frac{1}{N} \ll \frac{K}{{\cal E}_{J}} \ll N \Rightarrow
1 \ll R \ll N^{2}; \nonumber \\
{\rm Fock}&:& \ \ N \ll \frac{K}{{\cal E}_{J}} \Rightarrow
N^{2} \ll R.
\end{eqnarray}

Thus for the second-quantized version as for the first-quantum GP equation
version discussed above,
we have obtained a Hamiltonian with a form similar to those previously
derived, but with slightly different parameters, and with extra terms
that may be important for large atom-atom interactions.

\begin{figure}
\includegraphics[scale=.56]{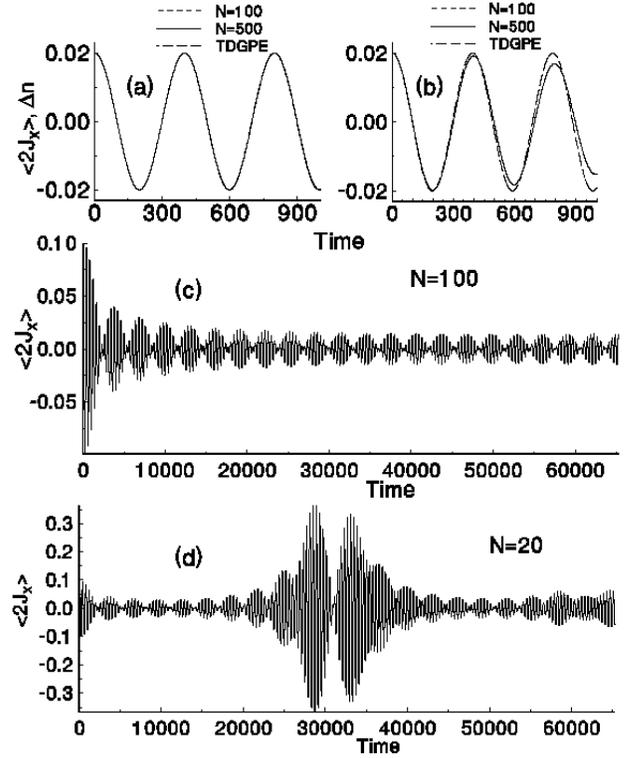}
\caption{In the limit of small $\bar{g}$ and small times, the temporal
behavior of $2\la J_x\ra$ and $\Delta n(t)$ approach each other. All of these
plots are for $\sigma = 1.5, V_{b}$=6.0.  (a) $\bar{g}$ =
1.0 $\times 10^{-4}$, $R$ = 0.017; (b-d) $\bar{g} = 1.0 \times 10^{-3}$,
$R$ = 0.17. For (a) and (b), the quantum and
classical periods are nearly the same, but (b) begins to show a
decrease of oscillation amplitude in the quantum case, due to the
second term in $H_{2}$. This modulation of the oscillations is shown over
longer time in (c), for $N$=100, and in (d), for $N$=20.}\label{qclass}
\end{figure}

Our formulation provides a connection with experimental conditions
through the Gross-Pitaevskii equation.
The expectation value $\la 2 J_x \ra$ describes the
difference between the number of particles in the two modes, and is
therefore an analog of the classical quantities, momentum, $z(t)$, and
number difference, $\Delta n(t)$.  The connection between $\langle 2J_{x}
\rangle$ and $\Delta n(t)$ can be most easily seen in the limit
of very small interactions (small $\bar{g}$), which is essentially the
``Rabi regime'' as defined in \cite{Leggett} and above.  The following
conclusions are based simply on an empirical evaluation of numerical
results.

\begin{figure}
\includegraphics[scale=.6]{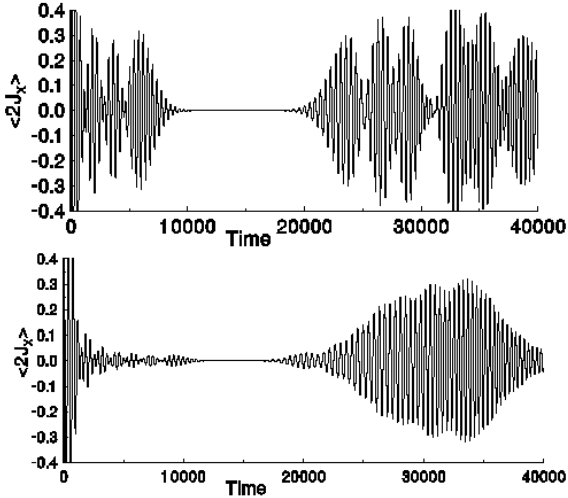}
\caption{Collapse and revivals for $\bar{g}$=3.0$\times 10^{-3}$, $R$=0.50,
for precisely defined $N=30$, but in (a) for precisely defined $N_{1}=27$ and
in (b) for a flat distribution $24 \leq N_{1} \leq 30$.}\label{wash}
\end{figure}

For $\bar{g} < 10^{-2}$, there
are clear tunneling oscillations with frequency $\Delta \beta$
from the first term in $H_{2}$ ($\Delta \gamma \ll \Delta \beta$ here).
These oscillations are modulated by effects from the second term
(in $J_{x}^{2}$) in $H_{2}$, which increase
with $\bar{g}$. For constant $\bar{g}=gN$, these modulations are independent
of $N$ over a large range of $N$, but undergo an additional modulation whose
period decreases with $N$, as shown in Fig. \ref{qclass}. This suggests that
there are various orders of time-dependent perturbations by which the
second term in $H_{2}$ perturbs the effect of the first.  However, we have
not been able to produce a quantitative perturbation-theoretic model.
For long enough
times, one observes the collapse and revival effects noted in \cite{Milburn}
and shown in Fig. \ref{wash}.  For larger values
of $\bar{g}$, these structures no longer persist.

Since it has been difficult to experimentally observe even a single
oscillation, and since it is difficult to control $N$ and the initial
imbalance, these oscillatory patterns may be impossible to observe. We
do find however, that the collapse and revival structure can persist
even if there is a spread of initial values of $N_{1} - N_{2}$.
Figure \ref{wash} compares results for fixed $N$=30, for an initial
sharp $N_{1}$ distribution, $N_{1}=27$, with results for an initial flat
distribution over the range $24 \leq N_{1} \leq 30$.
The stability of a part of the revival structure
may occur because two parameters are needed to characterize a point
on the sphere that is isomorphic to the $SU(2)$ algebra used above.
More extensive numerical results of phase space oscillations are
given in \cite{Links}, and detailed studies of averages in phase space
using the Husimi distribution have been presented in \cite{Mahmud2}.

\section{Calculations in 3D}

\subsection{General Formalism}

In comparing with experimental results, the transverse confinement enters.
In this study, we consider moderate
transverse confinement, not approaching the Tonks-Girardeau regime \cite{MO}.
We have extended the above methods to 3D as follows.
We write the TDGPE first in MKS units, denoted by overbars:
\begin{eqnarray} \label{tdgpe3D}
i \hbar\frac{\partial \bar{\psi}}{\partial \bar{t}} = \left[-
\frac{\hbar^{2}}{2m} \bar{\nabla}^{2}
 + \frac{m}{2}  \sum_{i} \omega_{i}^{2} \bar{x}_{i}^{2}
\right. \hspace*{3cm} \nonumber \\
\left. + V_{B} + g_{3D} |\bar{\psi}(\bar{\bf x},t)|^{2}\right]
\bar{\psi}(\bar{\bf x},t) \hspace*{1.5cm}
\end{eqnarray}
where $m$ is the atomic mass, and $g_{3D} = 4 \pi \hbar^{2}a_{3D}/m$,
$a_{3D}$ is the 3D scattering length, and
$\int d{\bf \bar{x}}|\bar{\psi}({\bf \bar{x}})|^{2} = N$.
The external potentials of
interest here will include a purely harmonic term as given above,
plus a barrier term as a function of $z$ that will
be chosen to be Gaussian or proportional to a $\cos^{2}$ function, as in
the experiments of \cite{Albiez}.

We let
\begin{eqnarray}
\omega_{x} = \omega_{y} = \eta \omega_{z},
\end{eqnarray}
and scale the coordinates and time as
\begin{eqnarray}
\bar{x}_{i} = \alpha_{i} x_{i}; \ \ \alpha_{i}^{2} = \hbar/m\omega_{i};
 \ \ \bar{t} = t/\omega_{z}.
\end{eqnarray}
Then since
\begin{eqnarray}
\int \! d{\bf x} |\psi(x)|^{2} \! = \!N = \!\!
\int \!\! d{\bf \bar{x}}|\bar{\psi}(\bar{\bf x})|^{2}
\!=\! \alpha_{x} \alpha_{y} \alpha_{z} \!\!
\int \! d{\bf x}|\bar{\psi}({\bf x})|^{2},
\end{eqnarray}
$\bar{\psi} = (\alpha_{x} \alpha_{y} \alpha_{z})^{-1/2} \psi$, and
(\ref{tdgpe3D}) becomes
\begin{eqnarray} \label{TDGP3D}
i \frac{\partial \psi(x,y,z,t)}{\partial t} =
{\cal H}(\lambda,\psi) \psi(x,y,z,t):
\hspace*{2.5cm}
\nonumber \\ {\cal H}(\lambda,\psi) =  - \frac{\eta}{2}\left(
\frac{\partial^{2}}{\partial x^{2}} + \frac{\partial^{2}}{\partial y^{2}}
\right) - \frac{1}{2} \frac{\partial^{2}}{\partial z^{2}}
+ \frac{\eta}{2} (x^{2} \! + \!y^{2}) \nonumber \\
+ \frac{z^{2}}{2} + V_{B} + 4 \pi \eta
\left(\frac{a_{3D}}{\alpha_{z}} \right) |\psi(x,y,z,t)|^{2} \hspace*{5mm}
\end{eqnarray}
where $\lambda$ represents the arguments $\eta,V_{B}, N, a_{3D}$ and
$\alpha_{z}$.

An ansatz analogous to (\ref{anz}) can now be introduced:
\begin{eqnarray}
\psi({\bf x},t) &=& \sqrt{N}[\psi_{1}(t) \Phi_{1}({\bf x}) + \psi_{2}(t)
\Phi_{2}({\bf x})];\label{anz3D}
\end{eqnarray}
\begin{eqnarray} \label{Phi123D}
\Phi_{1,2}({\bf x})& =& \frac{\Phi_{+}({\bf x})
\pm \Phi_{-}({\bf x})}{\sqrt{2}},
\end{eqnarray}
where
\begin{eqnarray}
\Phi_{\pm}(x,y,z) = \pm \Phi_{\pm}(x,y,-z); \ \ \int dxdy dz \Phi_{\pm}^{2} = 1.
\end{eqnarray}
The stationary GP equations take the form:
\begin{eqnarray} \label{modes3}
\beta_{\pm} \Phi_{\pm}(x,y,z) =  {\cal H}(\lambda,\Phi_{\pm})
\Phi_{\pm}(x,y,z).
\end{eqnarray}

Because the transverse wavefunction is very nearly Gaussian, some authors
have simply assumed a Gaussian, possibly with a $z$-dependent width,
and obtained modified equations \cite{Salasnich,Das} for what we have
called $\psi_{1}(z)$. Because we wanted to consider cases where the
Gaussian form may not be valid, we used general 3D algorithms.
Initial $\Phi_{\pm}$ wavefunctions were obtained by diagonalizing the DVR
Hamiltonian \cite{DVR1} using sparse matrix techniqes \cite{Davidson},
which made calculations with $>$120,000 mesh points possible
in minutes on a PC. To calculate the time evolution, the split-operator
method \cite{Feit} with Fast Fourier transform \cite{FF} was used, requiring
an hour or more of $\approx$2 GHz CPU time,
in view of the small time steps required.

From the $\Phi_{\pm}$ functions calculated from the 3D time-independent
Gross-Pitaevskii equation, one can also obtain the parameters $\beta_{\pm},
\gamma_{i,j}$, $A, B$ and $C$ as in Section \ref{Models1D}, to provide a
two-mode representation of tunneling oscillations in 3D.
In translating results from 1D to 3D for $g_{1D} = g_{3D} = 4 \pi
\eta a_{3D}/\alpha_{z}$, we find that, in the limit of weak interactions
and $\eta \geq 1$, the $\gamma_{ij}$ functions for 3D are a factor of
$2 \pi$ smaller than the corresponding 1D $\gamma_{ij}$ functions. The
explanation touches on the basic properties of tight transverse confinement.

\begin{figure}
\includegraphics[scale=.55]{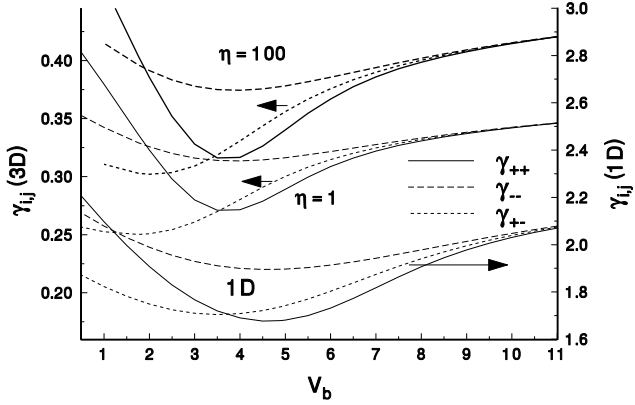}
\caption{Values for $\gamma_{ij}, i,j=+,-$ from 3D calculations with
$\eta$=1 and 100 (left axis), as compared with a 1D calculation (right axis).
Solid lines denote $\gamma_{++}$, short dashes $\gamma_{+-}$, and longer
dashes $\gamma_{--}$.  For 3D, the $\gamma_{ij}$ are approximately
$2\pi$ larger than for 1D.} \label{gam3D}
\end{figure}

If the transverse confinement is symmetric in $x$ and $y$ and is tight enough,
the 3D wavefunction $\Phi_{+}(x,y,z)$ can be factored into a function of $z$
and a function of $\rho = \sqrt{(x^2 + y^2)}$. Then if also the interactions
are weak enough, the $\rho$ function will be a Gaussian:
\begin{eqnarray}
\Phi_{+}(x,y,z) = \Phi_{+}(\rho,z) = e^{-\rho^{2}/2} \phi_{1}(z).
\end{eqnarray}
The normalization condition is
\begin{eqnarray}
1 = \int dx dy dz \Phi_{+}(x,y,z)^{2}  \hspace*{2.5cm} \nonumber \\
= 2 \pi \int d\rho \rho e^{-\rho^2} \int dz \phi_{1}(z)^{2}
= \pi \int dz \phi_{1}(z)^{2}.
\end{eqnarray}
Under the above conditions, and if $g_{1D} = g_{3D}$, then to within a
constant of proportionality, $\phi_{1}(z)$ will also be a solution of the
1D problem: $\phi_{1}(z) \propto \Phi_{+}^{1D}(z)$. For the 1D problem,
$\int dz |\Phi_{+}^{1D}(z)|^{2} = 1$, so from the different normalizations,
we see that, under all the above stated conditions,
\begin{eqnarray}
\phi_{1}(z) = \frac{1}{\sqrt{\pi}} \Phi_{+}^{1D}(z).
\end{eqnarray}
The 3D version of $\gamma_{++}$ becomes
\begin{eqnarray}
\gamma_{++}^{3D} = \int dx dy dz |\Phi_{+}^{3D}|^{4} \hspace*{2.5cm}
\nonumber \\
 =  2 \pi \int \! \rho d\rho e^{-2\rho^2} \int \! dz \phi_{1}(z)^{4}
 = 2 \pi \frac{1}{4} \frac{\gamma_{++}^{1D}}{\pi^{2}} =
\frac{\gamma_{++}^{1D}}{2 \pi}.
\end{eqnarray}
Similar relations hold also for $\gamma_{--}^{3D}$ and $\gamma_{+-}^{3D}$.
For larger interactions, the $\rho$ dependence is not exactly Gaussian,
the functions $\Phi_{\pm}^{3D}$ no longer factorize, and
the parameters $\gamma_{ij}^{3D}$ deviate from the above relations.
Figure \ref{gam3D} shows plots of $\gamma_{++}, \gamma_{--}$ and $\gamma_{+-}$
from 3D calculations with $\eta$ = 1 and 100, as compared with 1D results.
For $\eta$=100, the wavefunction is more concentrated than for $\eta=1$,
so the values for $\gamma_{ij}$ are slightly larger. Each is 5 to 6
times smaller than for the 3D case. Otherwise the dependences on $V_{b}$
are very similar.

\begin{figure}
\includegraphics[scale=.51]{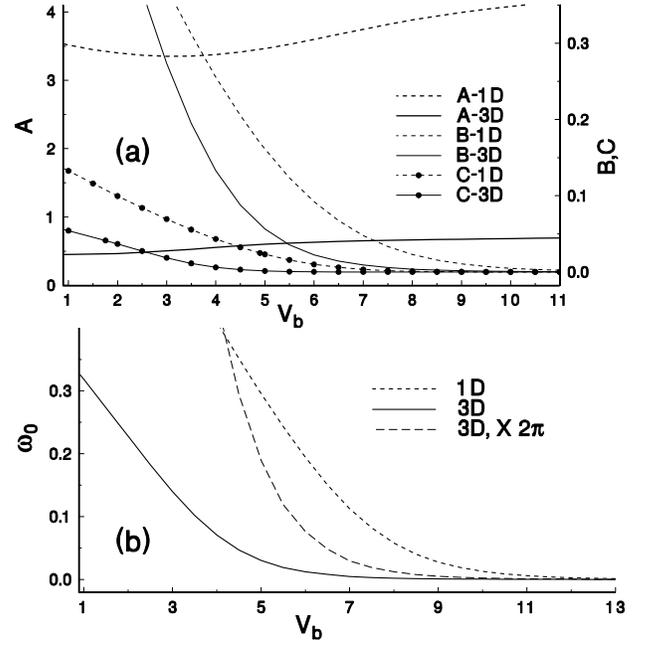}
\caption{(a) A comparison of calculated values for the parameters $A, B$ and
$C$ for 1D and for 3D, for $\eta = 1$ and $\bar{g}$=10. (b) A
comparison of Josephson plasma oscillation frequencies, $\omega_{0}$, for
1D and 3D, also for $\eta = 1$ and $\bar{g}$=10.} \label{ABC3D}
\end{figure}

There are other differences between 3D and 1D properties.
The difference energy, $\Delta \beta$, and hence also the parameter $B$
decrease more rapidly as a function of barrier height. Figure \ref{ABC3D}
compares the parameters $A, B$ and $C$ in 3D ($\eta=1$) and in
1D, for the case $\bar{g}=10$. Evidently, finite transverse confinement
decreases the difference between the symmetric and antisymmetric
condensate energies.  The differences are much the same for $\eta$ =100 as
for $\eta$=1. Also for $\eta = 1$ $\bar{g}$=10, Fig. \ref{ABC3D}b shows
that the plasma oscillation frequency in the limit of small $z, \phi$,
for barrier height, $V_{b} >5$, is even less than a
factor of $2 \pi$ smaller in 3D than in 1D.  This statement
has been found to be true for $\eta$=1 and 100, and $\bar{g}$ up to 10.

We conclude that two-mode models tend to be even more valid in 3D than
in 1D.

\begin{figure}
\includegraphics[scale=.55]{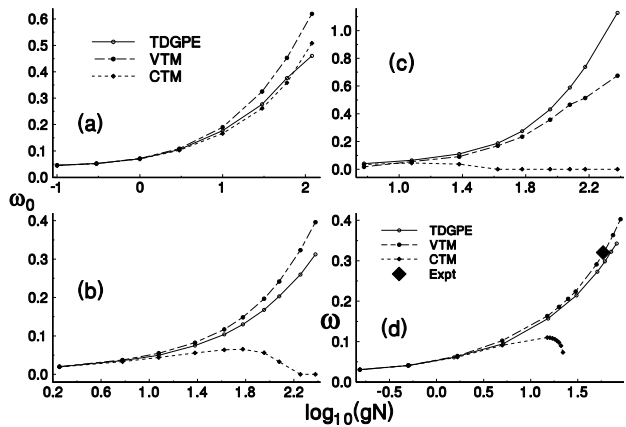}
\caption{Josephson plasma oscillation frequencies calculated from the
TDGPE and from VTM and CTM models, using (\ref{omegas}) and the
$\beta_{\pm}, \epsilon_{\pm}$ and $\gamma_{\pm}$ parameters obtained from
the 3D wavefunctions, $\Phi_{\pm}$.  For (a-c), Gaussian barriers were used,
as all the 1D calculations. For (d), a $\cos^{2}$ function to reproduce
the barrier in \cite{Albiez}. Other parameters were (a) $\eta$=1.0;
$V_{b}$=5.0; (b) $\eta = 1.0, V_{b}= 8.0$; (c) $\eta =100, V_{b} = 8.0$;
(d) Corresponding closely to the experiments in \cite{Albiez} (see text),
$\nu_{x}$ = 66 Hz,  $\nu_{y}$ = 90 Hz, $\nu_{z}$ = 78 Hz, barrier
$V_{b}$ = 5.28 $\hbar \omega_{z}$ times the cos$^{2}$ function given in
the text. The values plotted
are for $z_{0}=0.28$ as in the experiments, rather than for the limiting
case of small $z,\phi$, hence are labeled $\omega$ rather than $\omega_{0}$.}
\label{tper3D}
\end{figure}

Linear combinations of $\Phi_{\pm}$ functions provide the initial condition
for the TDGPE, for which we use the split operator approach with fast
Fourier transform \cite{Feit,FF}. To be able to compare TDGPE results with
(\ref{omegas}), we use a very small initial imbalance ($z_{0}$ =0.002) for
the TDGPE calculations. For the two-mode models, parameters are obtained
from wavefunctions calculated with the time-independent 3D GP equation,
as for 1D results above.  The results for $\omega_{o}$
are shown in Fig. \ref{tper3D}a-c. (Fig. \ref{tper3D}d pertains to the
experiments of \cite{Albiez} as discussed below). The plasma oscillation
frequency obtained from the TDGPE increases rapidly beyond
$\bar{g} = gN \approx$ 3. The two-mode model results match the TDGPE results
well for $\bar{g} < 1$.  Fig. \ref{tper3D}a, for Gaussian barrier of height
$V_{b}=5 \hbar \omega_{z}$, shows good agreement
for both the VTM and CTM with TDGPE results, up to $\bar{g}$ = 100.
On the other hand, when the barrier height is raised to 8$\hbar \omega_{z}$,
the CTM fails for $\bar{g} > 30$, for both $\eta=1$ (b) and $\eta=100$
(c). For the latter, the VTM result begins to deviate significantly from the
TDGPE value around $\bar{g}$=100. The failure of the CTM here is analogous
to the situation shown in Fig. \ref{omegao}, and occurs because ${\cal K}$
becomes negative.

\subsection{Comparison with Recent Experiments}

Very recently, the first quantitative experimental observations of
oscillations of Bose
condensates in a double well potential have been performed \cite{Albiez}.
The harmonic confinement was created by overlapping tightly focussed
Gaussian laser beams.  The harmonic frequencies were 66, 90 and 78 Hz
in what we will call the $x, y$ and $z$ directions. To produce the
double well, an optical standing wave from two beams of wavelength 811 nm, at
an angle of 9$^{o}$ were added, producing a barrier of the form
$V_{b} \cos(z\pi/w)^{2}$, with $V_{b} =$ 413(20) Hz, and $w$
= 5.20(20)$\mu$m.  1150 $^{87}$Rb atoms were loaded into this trap.
We have modeled this experimental configuration and find the effective
value of $gN=58.8$ from \ref{TDGP3D}, which corresponds to
$\bar{g} \approx 10$ in 1D simulations.

The reported experimental period of oscillation for $z_{0} = 0.28(6)$
was 40(2) msec. \cite{Albiez}, which corresponds to the value indicated
by the large diamond in Fig. (\ref{tper3D})d.  We obtain values very close
to this
observed tunneling frequency with both the VTM and TDGPE by using
a value for $a_{3D} = 100 a_{0}$ (where $a_{0}$ is the Bohr
radius) \cite{Verhaar}.  For this initial value $z_{0}$,
(although not for very small values of $z_{0}$),
self-trapping occurs with the CTM model, when based on solutions of the
Gross-Pitaevskii equation.  The calculated CTM frequency drops rapidly
before this point, as shown in Fig. (\ref{tper3D})d.

Note that by
reference to Figs. \ref{tper3D}a-c, we conclude that as long as the
temperature is sufficiently low (the temperature for
the experiments of \cite{Albiez} was immeasurably low), the aspect
ratio is not important, as results for $\eta=100$ and =1 are very similar,
but the relatively large value of the nonlinear term is important
in determing the validity of two-mode models.

Using the TDGPE, we have calculated a value of $z_{c}$=0.39 for the
stated conditions of these experiments, which is consistent with the observed
oscillations at $z_{0} = 0.28(6)$ and self-trapping for
$z(0) = 0.62(6)$, but lower than the value of $z_{c}$ = 0.50(5) quoted
in \cite{Albiez}. In this paper, the authors performed calculations with
the transverse Gaussian model of \cite{Salasnich} and obtained good
agreement with experimental observations. Our contribution is simply to
show that a two-mode model, with parameters from GP eigenfunctions,
also comes quite close to reproducing the experimental observations.

The other experiments that helped to motivate this study were performed
at NIST, MD, with $^{87}$Rb atoms in a ``pattern-loaded'' optical lattice.
The atoms were first loaded into a coarse lattice from Bragg-diffracted laser
beams, and then a finer lattice was turned on, such that every third lattice
site was occupied \cite{Peil}.  We are presently working to develop a
modification of the present approach to deal with such phenomena in a
periodic lattice.

\section{Conclusions}

By rigorous solution of coupled equations for the symmetric and anti-symmetric
wavefunctions for a Bose condensate in a double well potential, we have
derived equations for a new two-mode model that provides for variation of the
tunneling parameter with time, depending on the differences of number and phase
of atoms in the two wells.  We have compared results from this ``variable
tunneling model'' (VTM) with results from other two-mode models, from
multi-mode models that we have constructed, and from solutions of the
time-dependent Gross-Pitaevskii equation (TDGPE). In making these comparisons,
we numerically compute wavefunctions from the stationary Gross-Pitaevskii
equation and use appropriate integrals of these wavefunctions in the model
equations.  For small values of the non-linear
interaction term and moderate potential barriers, all the models agree nicely.
When the nonlinear interaction term exceeds a certain value, the tunneling
parameter in the usual ``constant tunneling model'' (CTM) becomes negative,
and thus the Josephson plasma oscillation frequency becomes imaginary.
The VTM remedies this problem, and produces better agreement with results
of the TDGPE.  We have performed such comparisons for 1D situations and
also for 3D situations, for which we have obtained 3D solutions of the
stationary and time-dependent GP equations.  The recent experimental
observations of tunneling oscillations and macroscopic self-trapping
of \cite{Albiez} are in the regime of moderately strong non-linear
interactions because of the large number of atoms (1150 $^{87}$Rb atoms).
Results from both the TDGPE and VTM for the observed tunnelling oscillation
frequency are in good agreement with the experimental value. However,
under the conditions of the experiment, the CTM when based on parameters
from the GP equation, leads to self-trapping rather than oscillation.

We also have applied our approach to derive an improved
Hamiltonian for quantum calculations, but find no reliable standards to
compare this approach with other quantization approaches. What we have
not investigated here are damping effects of thermal excitations, as
considered in \cite{Zapata} and \cite{Marino}.

We gratefully acknowledge support from NSF Grant PHY0354211, and
from the ONR.  We are especially indebted to M. Oberthaler for
sending a preprint of \cite{Albiez} before publication.
Communications with M. Olshanii, V. Dunjko, H. T. C. Stoof, V.
Korepin and D. Averin have been valuable in preparing this report.
In addition, we thank A. Muradyan for a critical reading of the
manuscript.

\end{document}